\documentclass[a4paper,aps,amsmath,amssymb,showpacs,10pt,twocolumn]{revtex4}   
\usepackage{epsfig}   
\usepackage{graphicx}   
\usepackage{amsmath,amssymb}

\newcommand{\bracket}[1]{\left\langle #1\right\rangle}   
   
\newcommand{\be}{\begin{equation}}   
\newcommand{\ee}{\end{equation}}   
\newcommand{\bd}{\begin{displaymath}}   
\newcommand{\ed}{\end{displaymath}}

\newcommand{\bsigma}{{\mbox{\boldmath $\sigma$}}}

\newcommand{\bmu}{{\mbox{\boldmath $\mu$}}}   
\newcommand{\bnu}{{\mbox{\boldmath $\nu$}}}

\newcommand{\bA}{\ensuremath{\mathbf{A}}}

\newcommand{\bU}{\ensuremath{\mathbf{U}}}

\newcommand{\btau}{{\mbox{\boldmath $\tau$}}}

\begin{document}   
\title{Cavity approach for real variables on diluted graphs and application to synchronization in small-world lattices}
\author{Nikos S Skantzos}\affiliation{Instituut voor  Theoretische Fysica, Celestijnenlaan 200D, Katholieke Universiteit Leuven B-3001, Belgium}
\email{Nikos.Skantzos@fys.kuleuven.be}
\author{Isaac \surname{P\'{e}rez Castillo}} \affiliation{Rudolf Peierls Centre for Theoretical Physics,University of Oxford, 1 Keble Road, Oxford, OX1 3NP, United Kingdom}
\email{isaac@thphys.ox.ac.uk}
\author{Jonathan P L Hatchett}\affiliation{Laboratory for Mathematical Neuroscience, RIKEN Brain Science Institute, Saitama 351-0198, Japan}
\email{hatchett@brain.riken.jp}
\begin{abstract}
We study XY spin systems on small world lattices for a variety of graph structures, e.g.\@ Poisson and scale-free, superimposed upon a one dimensional chain. In order to solve this model we extend the cavity method in the one pure-state approximation to deal with real-valued dynamical variables. We find that small-world architectures significantly enlarge the region in parameter space where synchronization occurs. We contrast the results of population dynamics performed on a truncated set of cavity fields with Monte Carlo simulations  and find excellent agreement. Further, we investigate the appearance of replica symmetry breaking in the spin-glass phase by  numerically analyzing the proliferation of pure states in the message passing equations.\\
\begin{center}
FoG 2005/04
\end{center}
\end{abstract}

\pacs{75.10.Nr, 05.20.-y, 64.60.Cn}
\maketitle

\section{Introduction}
Typically, given a model of an interacting spin system on a $d$-dimensional regular lattice, the first approach in understanding its properties is through the Weiss mean-field theory. Equivalently, one can neglect the underlying lattice structure completely, where each spin interacts with a finite number of particles, and assume that all spins interact with each other. In other words, one substitutes the original model by its (exactly solvable) fully-connected version. This crude step may fail either completely, for instance predicting phase transitions where there are none, or give a rather good qualitative picture of the physics of the model. In order to improve results both qualitatively and quantitatively one must then use sophisticated techniques, such as renormalization group.\\
During the last two decades, since the seminal paper of Viana and Bray \cite{Viana1985}, we have witnessed that it is still possible to consider interesting exactly solvable versions of the original model, other than the fully connected ones, while keeping the coordination number finite. Since then, interacting spin systems on random graphs have attracted quite some attention in spin glasses \cite{Viana1985, Kanter1987b,Mezard1987,Mezard2001,Mottishaw1987,Wong1988a}, neural networks \cite{Wemmenhove2003a,isaac,scalefree} and  small-world networks \cite{weigt,theodore}.\\
Surprisingly, during the last decade we have encountered interesting problems that can be mathematically formalized precisely as interacting spin systems on random graphs. Examples can be found in optimization problems \cite{Kirkpatrick1994,Monasson1997,Monasson1998b,Monasson1999}, error-correcting codes \cite{Murayama2000,ns1,Nishimori2001} and cryptography \cite{kms,ns2}.\\
In order to tackle this type of problems the replica approach and cavity method, fairly well understood for fully connected systems \cite{Mezard1987a}, had to be reconsidered \cite{Mezard1987,Mezard2001}: while in the one pure-state approximation \cite{Mezard1987a} a few order parameters suffice to describe the thermodynamics in the fully connected case, this no longer holds as soon as the coordination number is finite. In the latter case an infinite set of parameters, or a function order parameter, is required \cite{Monasson1998}. \\
Unfortunately, while most analytic studies of interacting spin systems have considered spins  of a discrete nature, \textit{e.g} Ising spins, much less attention has been given to systems with real-valued variables and in particular their model realization on random graphs.  Needless to say, such systems are not only more realistic from a fundamental physical point of view, but are also of special relevance in several contexts, as for instance in \@ Josephson junctions arrays \cite{wiesenfeld,granato,majhofer}, Kuramoto models \cite{kuramoto, kuramoto_review}, etc. \\
The goal of this paper is then twofold. Firstly, we wish to extend the cavity method to spin systems with real-valued variables. Secondly, and following the steps in \cite{softguzai}, we would like to study here the thermodynamical properties of Kuramoto models in small-world networks \cite{wattsstrogatz}. Small worlds, more than just curious structures, are apparently universal and can be observed in many different circumstances: from linguistic and social networks to the world-wide-web (for a recent review see e.g.\@ \cite{barabasi_review}). A large body of work has been devoted to the study of small-world networks, mainly numerically (see e.g.\@ \cite{watts_book}) and to a lesser extent analytically \cite{weigt,theodore}. \\
This paper is organized as follows: in Section \ref{sec:definitions} we introduce the XY spin model in small world networks. To solve it we generalize the cavity method in Section \ref{sec:CavityMethod} to real-valued dynamical variables to specific graph instances and in the ensemble case. We then use the cavity method on the XY small-world spin model in Section \ref{sec:XYmodel}: we first apply bifurcation analysis to calculate the continuous transition lines exactly. We also use population dynamics and compare the results with simulations. We finish Section \ref{sec:XYmodel} by studying the proliferation of pure states. The last section is for conclusions.
\section{Definitions}
\label{sec:definitions}
We study a small-world network of $N$ interacting XY spins $\bsigma_i=(\sigma^x_i,\sigma_i^y)$ with $i=1,\ldots,N$ and with $||\bsigma_i||^2=\bsigma_i^2=1$. At equilibrium the system interacts according to the Hamiltonian
\begin{equation}
\label{eq:hamiltonian}
H(\{\bsigma_i\}) = - J_0 \sum_i \bsigma_i\cdot\bsigma_{i+1} - J\sum_{(i,j)\in \mathcal{G}} \bsigma_i\cdot \bU_{ij}\, \bsigma_j\,,
\end{equation}
where the first term represents a ferromagnetic interaction along a one dimensional chain while the second one describes a chiral interaction. The factors $\bU_{ij}$ are orthogonal matrices drawn from some general distribution $D(\bU)$. Here $\mathcal{G}=(\mathcal{V},\mathcal{E})$ represents a graph with vertex set $\mathcal{V}=\{1,\ldots,N\}$ and edges $\mathcal{E}\ni(i,j)$, $\mathcal{E}\subseteq (\mathcal{V},\mathcal{V})$. We denote by $\partial i$ the neighbors of site $i$ and for any subset of vertices $A\subseteq \mathcal{V}$ we write $\sigma_A\equiv \{\sigma_i| i\in A\}$. 
\section{Generalized cavity method}
\label{sec:CavityMethod}
\subsection{Cavity method on an instance}
The cavity approach for diluted disordered systems with Ising-type variables has been developed in \cite{Mezard1987,Mezard2001}. The extension of the cavity method to any type of dynamical variable was worked out by one of us in \cite{IsaacThesis} and we would like to recall it here. To do so, let us first consider a general version of the Hamiltonian \eqref{eq:hamiltonian}, \textit{viz.} 
\begin{equation}
\label{eq:hamiltonian}
H(\{\bsigma_i\}) =  -J\sum_{(i,j)\in \widetilde{\mathcal{G}}} \bsigma_i\cdot \bU_{ij}\, \bsigma_j\,,
\end{equation}
where the nature of the spins is unspecified (Ising, real variables, spherical, multicomponent, and so on) and we will require only that the partition function is well-defined. Let us assume that the graph $\widetilde{\mathcal{G}}$ is tree-like (short loops are rare). It is then clear that the spins belonging to the neighborhood of $i$ are \textit{mainly} correlated through the spin at site $i$. If, however, we consider the fictitious situation of removing the spin at  site $i$, then in the resulting \textit{cavity graph} the joint probability distribution of its neighbors \textit{approximately} factorizes and we can write 
\begin{equation}
P^{(i)}(\bsigma_{\partial i})\overset{\text{Bethe}}{=}\prod_{j \in \partial i}P^{(i)}_j(\bsigma_j)\,,
\end{equation}
where $P^{(i)}_j(\bsigma_j)$ denotes the \textit{cavity probability distribution} of finding spin $j$ in state $\bsigma_j$ in the absence of site $i$. This is usually called the \textit{Bethe approximation}. Let us now consider the reverse situation where we start from a cavity graph without the spin at site $i$ and reincorporate the spin $i$ and reconnect all sites $\partial i$ to $i$. Then the \textit{physical} (\textit{marginal}) \textit{probability distribution} for $\bsigma_i$ on the original graph $\widetilde{\mathcal{G}}$ is given by
\begin{widetext}
\begin{equation}
\begin{split}
P_i(\bsigma_i)&=\frac{1}{Z_i} \int d \bsigma_{\partial i}\, e^{\beta J\sum_{\ell \in\partial i} \bsigma_i\cdot \bU_{i\ell}\, \bsigma_\ell}\,P^{(i)}(\bsigma_{\partial i})\overset{\text{Bethe}}{=}\frac{1}{Z_i} \int d \bsigma_{\partial i}\, e^{\beta J\sum_{\ell \in\partial i} \bsigma_i\cdot \bU_{i\ell}\, \bsigma_\ell}\,\prod_{\ell \in \partial i} P^{(i)}_\ell(\bsigma_\ell)\,,
\label{eq:true}
\end{split}
\end{equation}
\end{widetext}
with $Z_i=\int d\bsigma\, P_i(\bsigma)$. The marginal $P_i(\bsigma_i)$ is used to calculate the physical one-site quantities in the original graph $\widetilde{\mathcal{G}}$. Unfortunately the cavity distributions $P^{(i)}_\ell(\bsigma_\ell)$ are still not known. To overcome this problem we notice that that we get closed relations for them if we reconnect all sites but one $j\in\partial i$
\begin{widetext}
\begin{equation}
P^{(j)}_i(\bsigma_i)=\frac{1}{Z^{(j)}_i} \int d \bsigma_{\partial i\backslash j}\, e^{\beta J \sum_{\ell \in\partial i\backslash j} \bsigma_i\cdot \bU_{i\ell}\, \bsigma_\ell}\prod_{\ell \in \partial i\backslash j} P^{(i)}_\ell(\bsigma_\ell),\quad Z^{(j)}_i=\int d\bsigma\, P^{(j)}_i(\bsigma)\,.
\label{eq:cavityclose}
\end{equation}
\end{widetext}
We see now that eqs. \eqref{eq:cavityclose} provide recursion relations for the cavity distributions $P^{(j)}_i(\bsigma_i)$  $\forall i=1,\ldots,N$ and $\forall j\in\partial i$. Once these are known we can calculate any physical one-site quantity by using eqs. \eqref{eq:true}. For instance the magnetization vector $\textbf{M}$ and the spin-glass parameter matrix $\textbf{Q}$ are given by
\begin{equation}
\begin{split}
M_a&=\frac{1}{N}\sum_{i=1}^N\int d\bsigma\, P_i(\bsigma) \sigma_a\\
Q_{ab}&=\frac{1}{N}\sum_{i=1}^N\int d\bsigma d\btau \, P_i(\bsigma) P_i(\btau) \sigma_a \tau_b\,,
\label{sec:orderparameters}
\end{split}
\end{equation}
where $a,b$ labels the spin components. Note that the cavity distributions appearing in eqs. \eqref{eq:true} and \eqref{eq:cavityclose} are not expressed in terms of parameters. We will call this choice of representation of the cavity equations the \textit{parameter-independent representation}. \\
In contrast, let us go now to a \textit{parameter-dependent representation} of the cavity distributions. We denote a parametrization explicitly as $P_{i}^{(j)}(\bsigma)\to P(\bsigma|\bmu_{i}^{(j)})$. Since any probability distribution is defined to be positive the following exponential parametrization is convenient
\begin{equation}
P(\bsigma|\bmu_{i}^{(j)})=\exp\left(\sum_{n=0}^\infty \mu_{i,n}^{(j)}\varphi_n(\bsigma)\right)\,,
\label{eq:representation}
\end{equation}
where $\bmu_{i}^{(j)}=\{\mu_{i,n}^{(j)}\}_{n\geq0}$ are the so-called \textit{cavity fields} and $\{\varphi_n(\bsigma)\}$ is a set of orthogonal polynomials with respect to a weight $\omega(\bsigma)$
\begin{equation}
\int d\bsigma\, \omega(\bsigma)\varphi_n(\bsigma)\varphi_m(\bsigma)=\delta_{n,m}\,.
\end{equation}
We will assume that the first polynomial $\varphi_0(\bsigma)$ is a constant (the normalization constant of the probability). Now, using \eqref{eq:representation} in \eqref{eq:cavityclose} we find the following recursion relations for the cavity fields $\{\mu_{i,n}^{(j)}\}$
\begin{equation}
\begin{split}
 \mu_{i,m}^{(j)}&= \sum_{\ell \in \partial i\backslash j} u_m(\bmu_{\ell}^{(i)}|J,\bU_{i\ell}),\quad m>0\,,\\
u_m(\bmu|J,\bA)&\equiv\int d\bsigma\, w(\bsigma)\varphi_m(\bsigma)\\
&\times\log\int d \btau\, e^{\beta J \bsigma \cdot \bA\, \btau+\sum_{n>0}^\infty \mu_{n}\varphi_n(\btau)}\,,
\label{eq:cavity_fields}
\end{split}
\end{equation}
where $\textbf{u}(\bmu_{\ell}^{(i)}|J,\bU_{i\ell})=\{u_m(\bmu_{\ell}^{(i)}|J,\bU_{i\ell})\}_{m\geq 1}$ are the \textit{messages} or \textit{propagated fields}. Notice that if the dynamical variables are Ising-type, then only one cavity field suffices to parametrize the cavity distributions and we recover the method of \cite{Mezard1987,Mezard2001}.\\
Let us also use the same parametrization of the marginal distribution $P_i(\bsigma_i)\to P(\bsigma_i|\bmu_i)$, where the $\bmu_i$ are the \textit{physical fields}. Then from eqs. \eqref{eq:true} we get the following relation between the physical fields and the cavity fields
\begin{equation}
 \mu_{i,m}= \sum_{\ell \in \partial i} u_m(\bmu_{\ell}^{(i)}|J,\bU_{i\ell}),\quad m>0\,.
\label{eq:messagepassing}
\end{equation}
Equations \eqref{eq:cavity_fields} and \eqref{eq:messagepassing} are usually referred to as \textit{message passing equations} and have the following interesting physical meaning: when site $j\in \partial i$ is removed then each site $\ell\in\partial i\backslash j$ sends a message $\textbf{u}(\bmu_{\ell}^{(i)}|J,\bU_{i\ell})$ to site $i$ according to the interaction between spins $\ell$ and $i$ and the state of spin $\bsigma_\ell$. Then the spin $\bsigma_i$ finds its optimal configuration (through its cavity fields $\bmu_{i}^{(j)}$) in the cavity graph  by adding all messages up, eq. \eqref{eq:cavity_fields}.
\subsection{Cavity method on the ensemble}
Looking at the expressions of the order parameters we notice combinations like $(1/N)\sum_{i=1}^NP_i(\bsigma)$ $(1/N)\sum_{i=1}^NP_i(\bsigma) P_i(\btau)$ and so forth. It is therefore instructive to introduce a \textit{density of marginal distributions} $\mathcal{W}_{\text{phys}}[\{P\}]$ as follows
\begin{equation}
\begin{split}
\mathcal{W}_{\text{phys}}[\{P\}]=\frac{1}{N}\sum_{i=1}^N\delta_{(F)}\left[P(\bsigma)-P_i(\bsigma)\right]\,,
\label{eq:denstityphysicaldistribution}
\end{split}
\end{equation}
with $\delta_{(F)}(\cdots)$ a functional Dirac delta. Thus, we can rewrite the order parameters as follows
\begin{equation}
\begin{split}
M_a&=\int \{d P\}\mathcal{W}_{\text{phys}}[\{P\}]\int d\bsigma\, P(\bsigma) \sigma_a\\
Q_{ab}&=\int \{d P\}\mathcal{W}_{\text{phys}}[\{P\}] \int d\bsigma d\btau \, P(\bsigma) P(\btau) \sigma_a\tau_b\,.
\end{split}
\end{equation}
Analogously, we can also define a \textit{density of cavity distributions} $\mathcal{W}[\{P\}]$
\begin{equation}
\begin{split}
\mathcal{W}[\{P\}]&=\frac{1}{N}\sum_{i=1}^N\frac{1}{|\partial i|}\sum_{j\in \partial i}\delta_{(F)}\left[P(\bsigma)-P^{(j)}_{i}(\bsigma)\right]\,.
\label{eq:denstitycavitydistribution}
\end{split}
\end{equation}
If we then assume that the graph ensemble is characterized by a degree distribution $p_k$, from equations \eqref{eq:cavityclose} and \eqref{eq:denstitycavitydistribution} we obtain the following self-consistency equations for the density $\mathcal{W}[\{P\}]$ and by using equations \eqref{eq:true}, \eqref{eq:denstityphysicaldistribution}  and \eqref{eq:denstitycavitydistribution} we relate the densities  $\mathcal{W}_{\text{phys}}[\{P\}]$ and  $\mathcal{W}[\{P\}]$
\begin{widetext} 
\begin{eqnarray}
&&\hspace{-7mm}\mathcal{W}[\{P\}]=\sum_{k=0}^\infty \frac{p_k\,k}{\bracket{k}}\int\left[\prod_{\ell=1}^{k-1}\{dP_\ell\}\mathcal{W}[\{P_\ell\}] d \bU_\ell D(\bU_\ell) \right]\delta_{(F)}\left[P(\bsigma)-\frac{1}{Z}\int\left[\prod_{\ell = 1}^{k-1}\,d\btau_\ell P_\ell(\btau_\ell)\right]\, e^{J\beta  \sum_{\ell = 1}^{k-1}  \bsigma\cdot\bU_\ell\,\btau_\ell}\right]\label{eq:selfdensity1}\\
&&\hspace{-7mm}\mathcal{W}_{\text{phys}}[\{P\}]=\sum_{k=0}^\infty p_k\int\left[\prod_{\ell=1}^{k}\{dP_\ell\}\mathcal{W}[\{P_\ell\}] d \bU_\ell D(\bU_\ell) \right] \delta_{(F)}\left[P(\bsigma)-\frac{1}{Z}\int\left[\prod_{\ell = 1}^{k}\,d\btau_\ell P_\ell(\btau_\ell)\right]\, e^{J\beta  \sum_{\ell = 1}^{k}  \bsigma\cdot\bU_\ell\,\btau_\ell}\right]\label{eq:selfdensity2}\,.
\end{eqnarray}
\end{widetext}
Notice that the probability of choosing a cavity site  randomly  is proportional to the number of its bonds, i.e. $p_k k$, as reflected in the preceding equation. Finally we can also write self-consistency equations for the cavity fields $\bmu$ in the ensemble. We can do this in two ways. Let us start from  eq. \eqref{eq:cavity_fields} and analogously to the definition of the density of cavity distributions \eqref{eq:denstitycavitydistribution} we define a \textit{density of cavity fields}
\begin{equation}
w(\bmu)=\frac{1}{N}\sum_{i=1}^N\frac{1}{|\partial i|}\sum_{j\in \partial i}\delta\left(\bmu-\bmu^{(j)}_{i}\right)\,,
\label{eq:densitycavityfields}
\end{equation}
we then get
\begin{widetext}
\begin{equation}
\begin{split}
w(\bmu)&=\sum_{k=0}^\infty \frac{p_k\,k}{\bracket{k}}\int\left[\prod_{\ell=1}^{k-1}d \bmu_\ell w(\bmu_\ell) d \bU_\ell D(\bU_\ell) \right]\delta\left(\bmu- \sum_{\ell=1}^{k-1} \textbf{u}_m(\bmu_\ell|J,\bU_{\ell}) \right)\,.
\end{split}
\end{equation}
\end{widetext}
Alternatively, we would like to get the self-consistency equations for $w(\bmu)$  directly from those of $\mathcal{W}[\{P\}]$ as done in \cite{softguzai}. It is instructive to notice that we can get the transformations $w(\bmu)\leftrightarrow \mathcal{W}[\{P\}]$ from the definition \eqref{eq:densitycavityfields} and \eqref{eq:denstitycavitydistribution} of the density of cavity fields $w(\bmu)$ and cavity distributions $\mathcal{W}[\{P\}]$, respectively. Indeed, let us simply assume that for a given set of cavity fields $\bmu_{i}^{(j)}$ there is a unique cavity distribution  $P_{i}^{(j)}(\bsigma)$ and vice versa so that we are entitled to write $\bmu_{i}^{(j)}=\bmu(\{P_i^{(j)}\})$. Then from eqs. \eqref{eq:densitycavityfields} and  \eqref{eq:denstitycavitydistribution} we obtain
\begin{widetext}
\begin{equation}
\begin{split}
w(\bmu)&=\frac{1}{N}\sum_{i=1}^N\frac{1}{|\partial i|}\sum_{j\in \partial i}\delta\left(\bmu-\bmu(\{P_i^{(j)}\})\right)=\int\{dP\}\mathcal{W}[\{P\}]\delta\left[\bmu-\bmu(\{P\})\right]\\
\mathcal{W}[\{P\}]&=\frac{1}{N}\sum_{i=1}^N\frac{1}{|\partial i|}\sum_{j\in \partial i}\delta_{(F)}\left[P(\bsigma)-P^{(j)}_{i}(\bsigma|\bmu_{i}^{(j)})\right]=\int d\bmu w(\bmu)\delta_{(F)}\left[P(\bsigma)-P(\bsigma|\bmu)\right]\,,
\label{eq:transformation}
\end{split}
\end{equation}
\end{widetext}
respectively, which is precisely the transformation $w(\bmu)\leftrightarrow \mathcal{W}[\{P\}]$ for switching between parameter-dependent and parameter-independent representations used in \cite{softguzai}.\\
Finally the order parameters can be expressed as an integral over the density of cavity fields by using the transformations \eqref{eq:transformation}
\begin{equation}
\begin{split}
M_a&=\int d\bmu\, w_{\text{phys}}(\bmu)\int d\bsigma\, P(\bsigma|\bmu) \sigma_a\\
Q_{ab}&=\int d\bmu\, w_{\text{phys}}(\bmu) \int d\bsigma d\btau \, P(\bsigma|\bmu) P(\btau|\bmu) \sigma_{a}\tau_b\,,
\end{split}
\end{equation}
where $ w_{\text{phys}}(\bmu)$ is the density of physical fields.
\section{Application to the XY spin system in small worlds}
\label{sec:XYmodel}
Let us now consider the XY model in small worlds defined by the Hamiltonian \eqref{eq:hamiltonian}. The application of the cavity method is straightforward. We must take into account that the spins are XY, so that the integral over spin degrees of freedom is over the unit circle, $\int d\bsigma\to\int_{S} d\bsigma$ with $S=\{(\sigma_x,\sigma_y)\in\mathbb{R}^2:||\bsigma||=1\}$, and that the energetic contribution coming from the chain and from the graph is different.\\
Indeed, let us define $\mathcal{C}(i)=\{i-1,i+1\}$. We will distinguish whenever the neighbor $j\in\partial i$ belongs to the immediate neighbors of the chain, $j\in \mathcal{C}(i)$, or not, $j\in\partial i\backslash \mathcal{C}(i)$. Then messages transmitted along the chain between two spins $\bsigma$ and $\btau$ have an energetic contribution of $-J_0\bsigma\cdot\btau$ while for those messages along the graph the contribution is $-J\bsigma\cdot\bU\,\btau$. Thus the message passing equations in this case simply read
\begin{equation}
\begin{split}
 \mu_{i,m}^{(j)}&=\sum_{\ell \in \mathcal{C}(i)} u_m(\bmu_{\ell}^{(i)}|J_0,\openone_{i\ell})\\
&+\sum_{\ell \in \partial i\backslash j} u_m(\bmu_{\ell}^{(i)}|J,\bU_{i\ell}),\quad j\in\partial i\backslash \mathcal{C}(i)\\
 \mu_{i,m}^{(j)}&=\sum_{\ell \in \mathcal{C}(i)\backslash j} u_m(\bmu_{\ell}^{(i)}|J_0,\openone_{i\ell})\\
&+ \sum_{\ell \in \partial i} u_m(\bmu_{\ell}^{(i)}|J,\bU_{i\ell}),\quad j\in \mathcal{C}(i)\,,\\
\label{eq:instanceequations}
\end{split}
\end{equation}
where $\openone$  is the $2\times 2$ identity matrix. For the physical fields we get
\begin{equation}
\begin{split}
\mu_{i,m}&=\sum_{\ell \in \mathcal{C}(i)} u_m(\bmu_{\ell}^{(i)}|J_0,\openone_{i\ell})+ \sum_{\ell \in \partial i} u_m(\bmu_{\ell}^{(i)}|J,\bU_{i\ell})\,.
\end{split}
\end{equation}
In order to have the equations in the ensemble we define two densities of cavities fields, similar to definition \eqref{eq:densitycavityfields}, as follows
\begin{equation}
\begin{split}
w_{\text{gr}}(\bmu)&=\frac{1}{N}\sum_{i=1}^N\frac{1}{|\partial i\backslash \mathcal{C}(i)|}\sum_{j\in \partial i\backslash \mathcal{C}(i)}\delta\left(\bmu-\bmu^{(j)}_{i}\right)\\
 w_{\text{ch}}(\bmu)&=\frac{1}{N}\sum_{i=1}^N\frac{1}{|\mathcal{C}(i)|}\sum_{j\in \mathcal{C}(i)}\delta\left(\bmu-\bmu^{(j)}_{i}\right)\,,
\end{split}
\end{equation}
which lead us to the self-consistency equations
\begin{widetext}
\begin{equation}
\begin{split}
w_{\text{gr}}(\bmu)&= \int\left[\prod_{\ell=1}^2 d\bnu_\ell w_{\text{ch}}(\bnu_\ell)\right]\sum_{k\geq 0}\frac{p_k k}{\bracket{k}} \int\left[\prod_{\ell=1}^{k-1}d\bmu_\ell w_{\text{gr}}(\bmu_\ell)\,d \bU_\ell D(\bU_\ell)\right]\delta\left(\bmu-\sum_{\ell=1}^2\textbf{u}(\bnu_\ell|J_0,\openone)-\sum_{\ell=1}^{k-1}\textbf{u}(\bmu_\ell|J,\bU_\ell) \right)\\
w_{\text{ch}}(\bmu)&= \int d\bnu w_{\text{ch}}(\bnu)\sum_{k\geq 0} p_k  \int\left[\prod_{\ell=1}^{k}d\bmu_\ell w_{\text{gr}}(\bmu_\ell)\,d \bU_\ell D(\bU_\ell)\right]\delta\left(\bmu-\textbf{u}(\bnu|J_0,\openone)-\sum_{\ell=1}^{k}\textbf{u}(\bmu_\ell|J,\bU_\ell) \right)\,.
\label{eq:XYdensitiespop}
\end{split}
\end{equation}
\end{widetext}
Similarly, we get the following equation for the density of physical fields
\begin{widetext}
\begin{equation}
w_{\text{phys}}(\bmu)= \int\left[\prod_{\ell=1}^2 d\bnu_\ell w_{\text{ch}}(\bnu_\ell)\right]\sum_{k\geq 0}p_k  \int\left[\prod_{\ell=1}^{k}d\bmu_\ell w_{\text{gr}}(\bmu_\ell)\,d \bU_\ell D(\bU_\ell)\right]\delta\left(\bmu-\sum_{\ell=1}^2\textbf{u}(\bnu_\ell|J_0,\openone)-\sum_{\ell=1}^{k}\textbf{u}(\bmu_\ell|J,\bU_\ell) \right)\,.
\end{equation}
\end{widetext}
\subsection{Parameter-independent equations and bifurcation analysis}
The above set of self-consistency eqs. \eqref{eq:XYdensitiespop} do not allow a general analytical treatment. However, we can extract some information exactly. To do so, it is convenient to switch to the parameter-independent representation. After using the transformations \eqref{eq:transformation} we obtain
\begin{widetext}
\begin{equation}
\begin{split}
\mathcal{W}_{\text{gr}}[\{P\}]&=\sum_{k=0}^\infty \frac{p_k\,k}{\bracket{k}}\int \left[\prod_{\ell=1}^2\{dP^\ell_{\text{ch}}\}\mathcal{W}_{\text{ch}}[\{P^\ell_{\text{ch}}\}]\right] \int\left[\prod_{\ell=1}^{k-1}\{dP^\ell_{\text{gr}}\}\mathcal{W}_{\text{gr}}[\{P^\ell_{\text{gr}}\}] d \bU_\ell D(\bU_\ell)\right]\\
&\times \delta_{(F)}\left[P(\bsigma)-\frac{1}{Z_{\text{gr}}}\int_{S} \left[\prod_{\ell=1}^2 d\bsigma_\ell P^\ell_{\text{ch}}(\bsigma_\ell)\right]\int_S\left[\prod_{\ell = 1}^{k-1}\,d\btau_\ell P^\ell_{\text{gr}}(\btau_\ell)\right]\, \exp\left(\beta J_0 \sum_{\ell=1}^2\bsigma\cdot\bsigma_\ell+ \beta J \sum_{\ell = 1}^{k-1} \bsigma\cdot\bU_\ell\,\btau_\ell\right)\right]\\
\mathcal{W}_{\text{ch}}[\{P\}]&=\sum_{k=0}^\infty p_k\int \{dP_{\text{ch}}\}\mathcal{W}_{\text{ch}}[\{P_{\text{ch}}\}] \int\left[\prod_{\ell=1}^{k}\{dP^\ell_{\text{gr}}\}\mathcal{W}_{\text{gr}}[\{P^\ell_{\text{gr}}\}] d \bU_\ell D(\bU_\ell)\right]\\
&\times \delta_{(F)}\left[P(\bsigma)-\frac{1}{Z_{\text{ch}}}\int_S d\bsigma' P_{\text{ch}}(\bsigma')\int_S\left[\prod_{\ell=1}^{k}\,d\btau_\ell P^\ell_{\text{gr}}(\btau_\ell)\right]\, \exp\left(\beta J_0 \bsigma\cdot\bsigma'+ \beta J \sum_{\ell = 1}^{k} \bsigma\cdot\bU_\ell\,\btau_\ell\right)\right]
\label{eq:XYensemble}
\end{split}
\end{equation}
\end{widetext}
and similarly for the density of physical distributions
\begin{widetext}
\begin{equation}
\begin{split}
\mathcal{W}_{\text{phys}}[\{P\}]&=\sum_{k=0}^\infty p_k\int \left[\prod_{\ell=1}^2\{dP^\ell_{\text{ch}}\}\mathcal{W}_{\text{ch}}[\{P^\ell_{\text{ch}}\}]\right] \int\left[\prod_{\ell=1}^{k}\{dP^\ell_{\text{gr}}\}\mathcal{W}_{\text{gr}}[\{P^\ell_{\text{gr}}\}] d \bU_\ell D(\bU_\ell)\right]\\
&\times \delta_{(F)}\left[P(\bsigma)-\frac{1}{Z}\int_S \left[\prod_{\ell=1}^2 d\bsigma_\ell P^\ell_{\text{ch}}(\bsigma_\ell)\right]\int_S\left[\prod_{\ell = 1}^{k}\,d\btau_\ell P^\ell_{\text{gr}}(\btau_\ell)\right]\, \exp\left(\beta J_0 \sum_{\ell=1}^2\bsigma\cdot\bsigma_\ell+ \beta J \sum_{\ell = 1}^{k} \bsigma\cdot\bU_\ell\,\btau_\ell\right)\right]\,.
\label{eq:XYensemblephysical}
\end{split}
\end{equation}
\end{widetext}
Even though eqs. \eqref{eq:XYensemble} and \eqref{eq:XYensemblephysical} still look rather nontrivial they allow us to locate the continuous transitions in a clean way. Let us start by noticing that from the definitions of the magnetization and the spin-glass matrix overlap, the paramagnetic solution ($M_{a}=Q_{ab}=0$) is simply given by
\begin{equation}
\begin{split}
\mathcal{W}_{\text{phys}}[\{P\}]&=\delta_{(F)}\left[P(\bsigma)-\frac{1}{2\pi}\right]\,,
\label{eq:paramagn}
\end{split}
\end{equation}
which implies
\begin{equation}
\begin{split}
\mathcal{W}_{\text{gr}/\text{ch}}[\{P\}]=\delta_{(F)}\left[P(\bsigma)-\frac{1}{2\pi}\right]\,.
\label{eq:paramagn}
\end{split}
\end{equation}
To find bifurcations away from \eqref{eq:paramagn} we follow \cite{softguzai} and make the so-called Guzai expansion $P(\bsigma) \rightarrow (2\pi)^{-1} + \Delta(\bsigma)$, with $\mathcal{W}_{\text{gr}/\text{ch}}[\{P\}] \to \widetilde{W}_{\text{gr}/\text{ch}}[\{\Delta\}]$ and where normalization requires that $\widetilde{W}_{\text{gr}/\text{ch}}[\{\Delta\}] = 0$ if $\int_S d\bsigma\, \Delta(\bsigma) \neq 0$. We then examine quantities such as $\int_S \{d\Delta\}  \widetilde{W}_{\text{gr}/\text{ch}}[\{\Delta\}] \Delta(\bsigma_1)\cdots\Delta(\bsigma_r)$ for $r=1,2$, and assume that close to a continuous phase transition they are of order $\mathcal{O}(\epsilon^r)$ with $\epsilon$ arbitrarily small. Inspecting the possible bifurcations in first and second order in $\epsilon$ we find the following two transition lines written in terms of the moments $\bracket{k} $ and $\bracket{k^2}$ of the graph degree distribution $p_k$ (for details see similar calculations in \cite{softguzai}):
\begin{widetext}
\begin{eqnarray}
\text{P-F:}&&\quad 1 = \cos(\overline{\omega})\,\frac{I_1(\beta J)}{I_0(\beta J)}\, \left[\frac{\bracket{k^2}-\bracket{k}}{\bracket{k}}+2\bracket{k} \frac{I_1(\beta J_0)}{I_0(\beta J_0) -I_1(\beta J_0)} \right]\label{eq:PF}\\
\text{P-SG:}&&\quad 1=\frac{I_1^2(\beta J)}{I_0^2(\beta J)}\left[\frac{\bracket{k^2}-\bracket{k}}{\bracket{k}}+2\bracket{k}\frac{I_1^2(\beta J_0)}{I_0^2(\beta J_0)-I_1^2(\beta J_0)}\right]\,,
\end{eqnarray}
\end{widetext}
where we have used the following measure for $D(\bU)$
\begin{equation}
\begin{split}
D(\bU)&=\int d\omega \,K(\omega)\,\delta\left[\bU-\begin{pmatrix}\cos\omega&\sin\omega\\-\sin\omega&\cos\omega\end{pmatrix}\right]\,,\\
\end{split}
\end{equation}
with
\begin{equation}
\begin{split}
K(\omega)&=\frac{1}{2}\delta[\omega-\overline{\omega}]+\frac{1}{2}\delta[\omega+\overline{\omega}]
\end{split}
\end{equation}
for some $\overline{\omega}\in[0,2\pi]$. 
\subsection{Parameter-dependent equations, population dynamics and simulations}
\begin{figure}[h]
\setlength{\unitlength}{0.1cm}
\begin{picture}(100,73)
\put(15,5){\includegraphics[height=5.6cm,width=5.9cm]{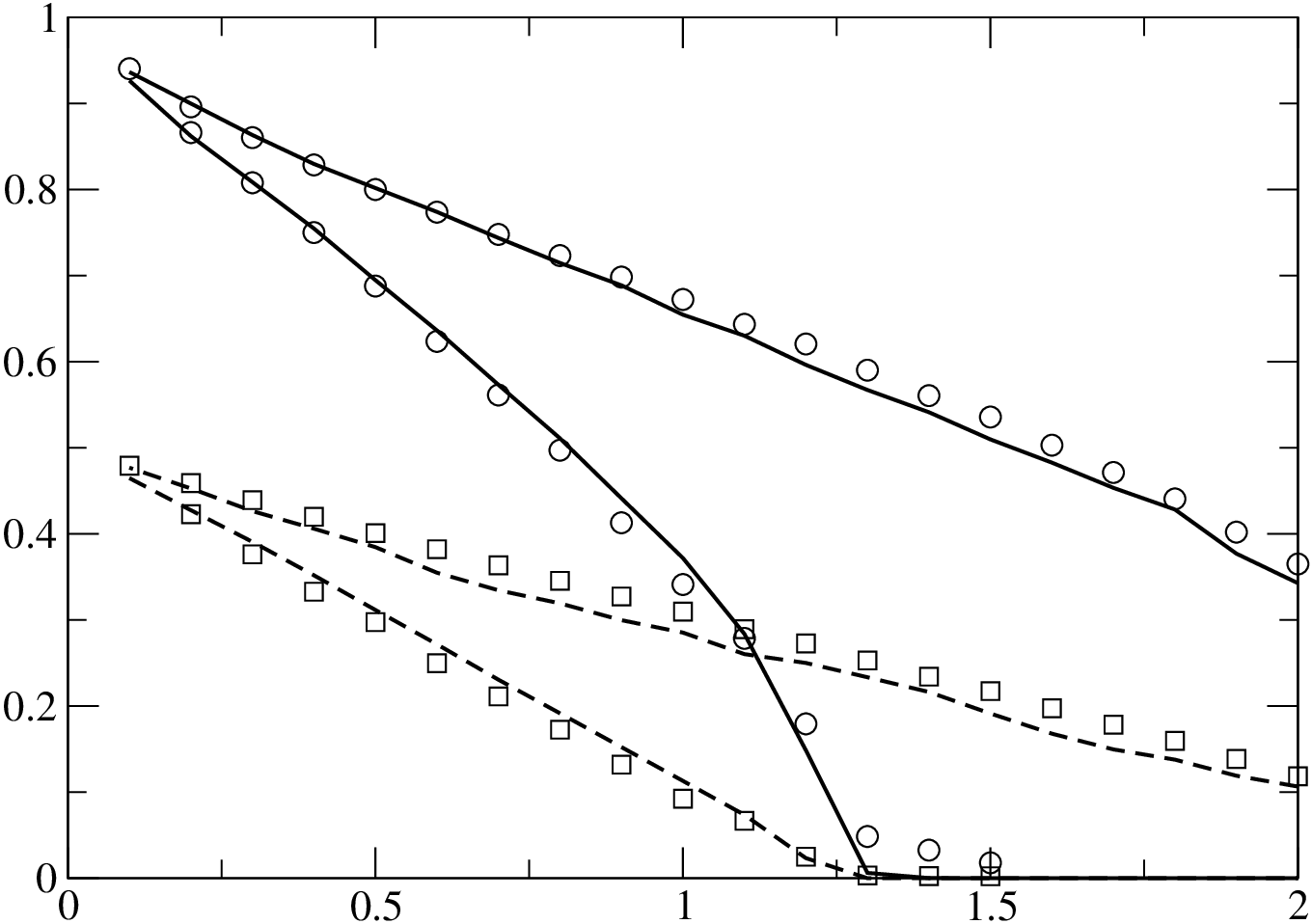}}
\put(06,28){$m,q$}
\put(45,-1){$c^{-1}$}
\end{picture}
\caption{Comparison between population dynamics and Monte Carlo simulations for the XY small-world systems with Poisson degree distribution and angular disorder $K(\omega)$ with $\overline{\omega}=\frac14\pi$. We plot the magnetization $m=\sqrt{M_x^2+M_y^2}$  (solid lines) and overlap $q=\frac12(Q_{xx}+Q_{yy})$ (dashed lines) against the inverse connectivity $c^{-1}$. Top lines refer to $T=0.3$ while lower ones to $T=0.5$. Markers correspond to simulation results of system size $N=40,000$. The critical parameter values are in good agreement with those obtained from our bifurcation analysis.}
\label{fig:population_dynamics}
\end{figure}

To obtain the behavior of the order parameters with respect to the control parameters (such as temperature, average connectivity and so on) we must solve the self-consistency equations numerically. In order to do so, we switch to the parameter-dependent representation for self-consistency equations \eqref{eq:XYensemble} and apply population dynamics.\\
To do the integrals in \eqref{eq:XYdensitiespop} it is convenient to parametrize the spin vector $\bsigma=(\cos(\phi),\sin(\phi))$ with $\phi\in[0,2\pi)$. Then for the cavity and physical distribution we have that $P(\bsigma)\to P(\phi)$ and since $P(\phi)$ is a periodic function of $\phi$ we choose a representation in Fourier modes
\begin{equation}
\begin{split}
P(\phi|\bmu)\sim\exp\left(\sum_{s\geq1} a_{s}\cos(s\phi)+\sum_{s\geq1} b_{s}\sin(s\phi)\right)\,,
\label{eq:fouriermodes}
\end{split}
\end{equation}
with $\bmu=(a_1,a_2,\ldots;b_1,b_2,\ldots)$, which implies that we have chosen, rather than polynomials $\varphi_n(\bsigma)$, plane-waves $\varphi_n(\bsigma)\to(\sin(n\phi),\cos(n\phi))$ with weight $w(\bsigma)=1/\pi$. After all this has been introduced into \eqref{eq:XYdensitiespop} we use population dynamics to solve them. Needless to say that to solve the equations for an infinite number of cavity fields is unfeasible: we must truncate the Fourier series in  the parameter representation \eqref{eq:fouriermodes}. Also, we describe the macroscopic state of the system in terms of just two order parameters $m = \sqrt{M_x^2 + M_y^2}$ and $q= (1/2)(Q_{xx} + Q_{yy})$. Numerical solutions of the eqs. \eqref{eq:XYdensitiespop} using population dynamics considering only two parameters $\bmu=(a_1,0,\ldots;b_1,0,\ldots)$  are shown in Fig. \ref{fig:population_dynamics}. To see the crudeness of the truncation we have compared the numerical results with Monte Carlo simulations. These have been done for system sizes of $N=40\cdot 10^3$ spins and  thermalizing with the Fast Linear algorithm \cite{Loison2004}. In Fig. \ref{fig:population_dynamics} the comparison between population dynamics and Monte Carlo simulations is fairly good ($J_0=1/2$ and $J=1$), supporting the truncation scheme. 
\subsection{Equations for a given instance and proliferation of pure states}
\label{sec:AT}
So far we have been tacitly assuming that there is only one pure state, or equivalently  one set of cavity fields $\bmu_{i}^{(j)}$ $\forall i=1,\ldots,N$ and $\forall j\in\partial i$, solving the message passing equations \eqref{eq:instanceequations}. Thus iteration of these equations does not depend on initial conditions. The proliferation of solutions to the equations \eqref{eq:instanceequations} signals ergodicity breaking.\\
To examine ergodicity breaking we follow a method similar to the one presented in \cite{pagnani}. For a given graph realization we take two different populations of cavity fields $\bmu^{(j),1}_i$ and $\bmu^{(j),2}_i$ and let them evolve according to the message passing equations \eqref{eq:instanceequations}.  If there were a single pure state then the system would be ergodic, and so after some transient, the state of the system would be independent of initial conditions. On the other hand, dependence on initial conditions for long times suggests that there are long range correlations contradicting the assumption of a single pure state (i.e. the clustering property is contravened) \cite{SteinNewman}. The difference between the two populations of fields is measured by introducing the following parameter
\begin{equation}
\lambda=\frac{1}{N}\sum_{i=1}^{N}\left(||\textbf{m}^{(1)}_i||-||\textbf{m}^{(2)}_i||\right)\,,
\label{eq:lambda_AT}
\end{equation}
where $\textbf{m}^{(1,2)}_i$  is the local magnetization for the two copies of the system. Thus, if $\lambda \neq 0$ the solution of the iterative message passing equations \eqref{eq:instanceequations} is non-ergodic. Notice that $\lambda$ is invariant under global rotations between the two copies and therefore does not take into account the rotational degeneracy of the Hamiltonian.
\subsection{Phase diagrams}
\begin{figure}[t]
\setlength{\unitlength}{0.1cm}
\begin{picture}(70,150)
\put(3,110){\includegraphics[height=4.cm,width=6.6cm]{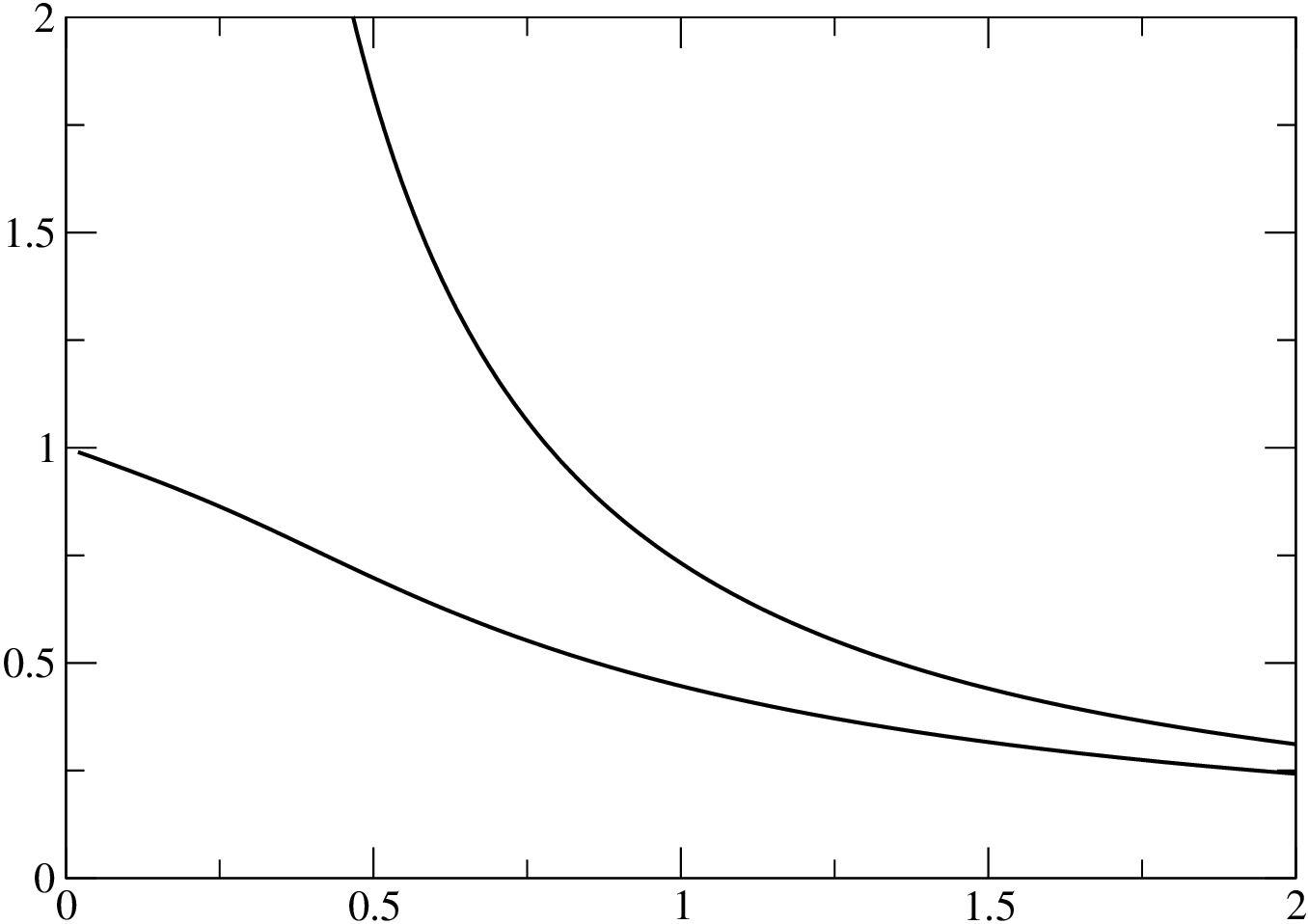}}
\put(3,60){\includegraphics[height=4.cm,width=6.6cm]{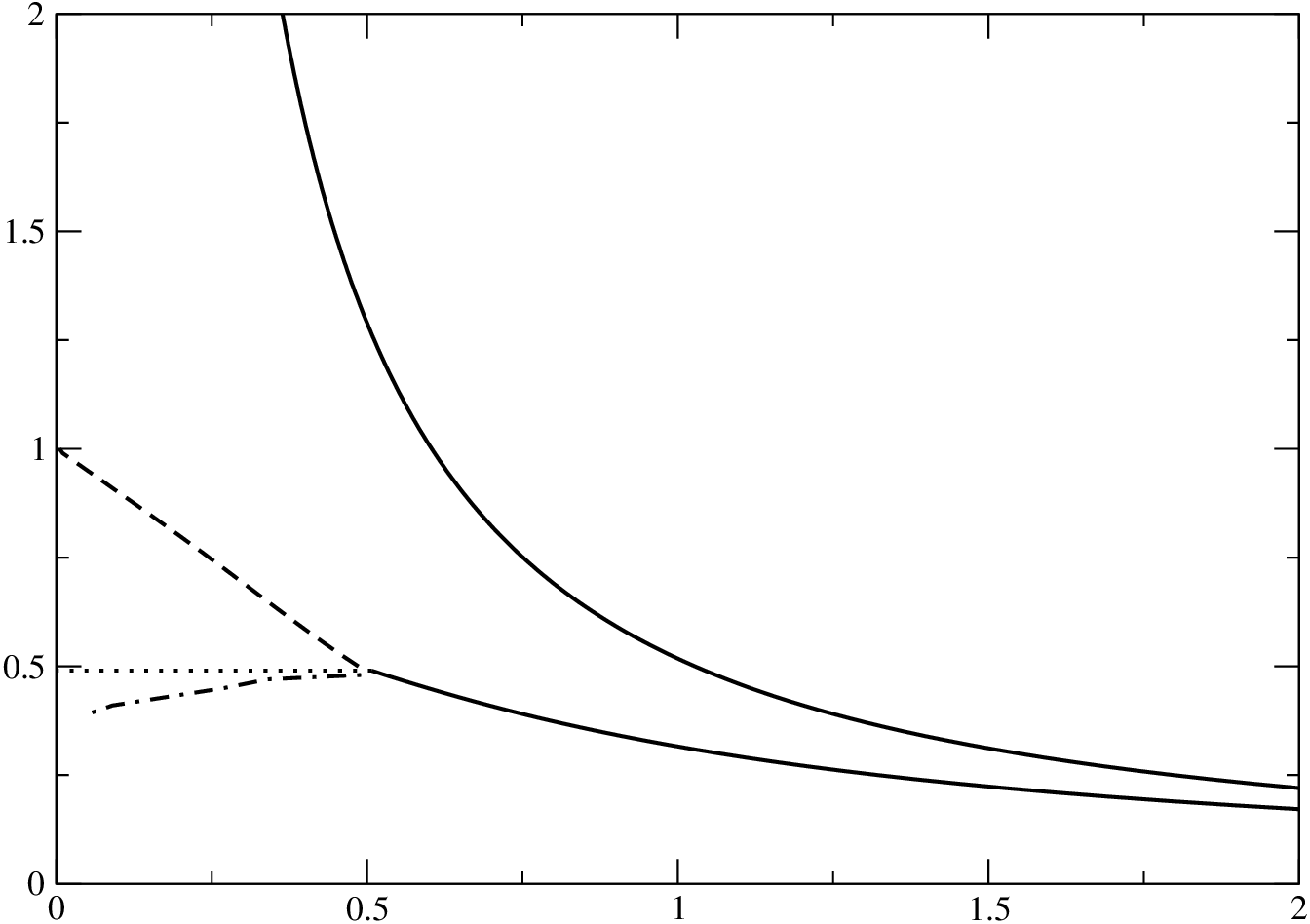}}
\put(3,10){\includegraphics[height=4.cm,width=6.6cm]{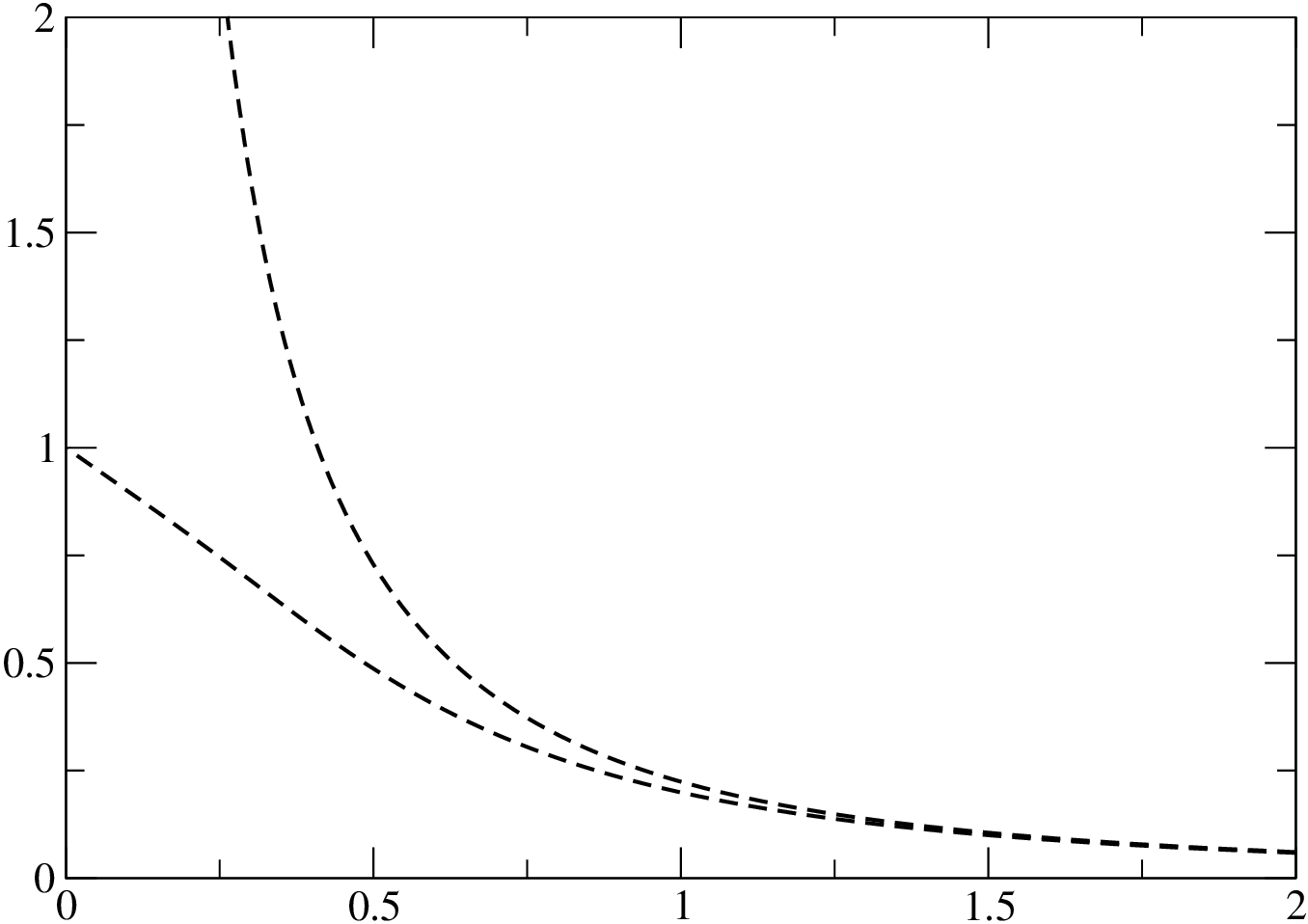}}
\put(35,5){$T/J$}
\put(-2,30){$c^{-1}$}\put(-2,80){$c^{-1}$}\put(-2,130){$c^{-1}$}
\put(40,135){P}\put(20,120){F}
\put(20,65){F}\put(40,75){P}\put(40,30){P}
\put(15,15){SG}
\end{picture}
\caption{Phase diagrams for XY small-world networks with bimodal distribution $K(\omega)=\frac12\delta[\omega-\overline{\omega}]+\frac12\delta[\omega+\overline{\omega}]$ on Poisson graphs for which $p_k=c^k\,e^{-c}/k!$ and $\overline{\omega}=0$ (top), $\overline{\omega}=\frac14\pi$ (middle) and $\overline{\omega}=\frac12\pi$ (bottom). In all diagrams, solid lines represent the P$\to$F transition whereas dashed ones represent P$\to$ SG ones (lower lines: $J_0=0$ and upper: $J_0=0.5$). In the middle panel, where for $J_0=0$ all three phases meet, dotted lines represent the F$\to$SG transition and dot-dashed the boundary above which $\lambda>0$ (\ref{eq:lambda_AT}) and non-trivial ergodicity breaking takes place. }
\label{fig:phase_bimodal}
\end{figure}
\begin{figure}[h]
\setlength{\unitlength}{0.1cm}
\begin{picture}(70,150)
\put(3,110){\includegraphics[height=4.cm,width=6.6cm]{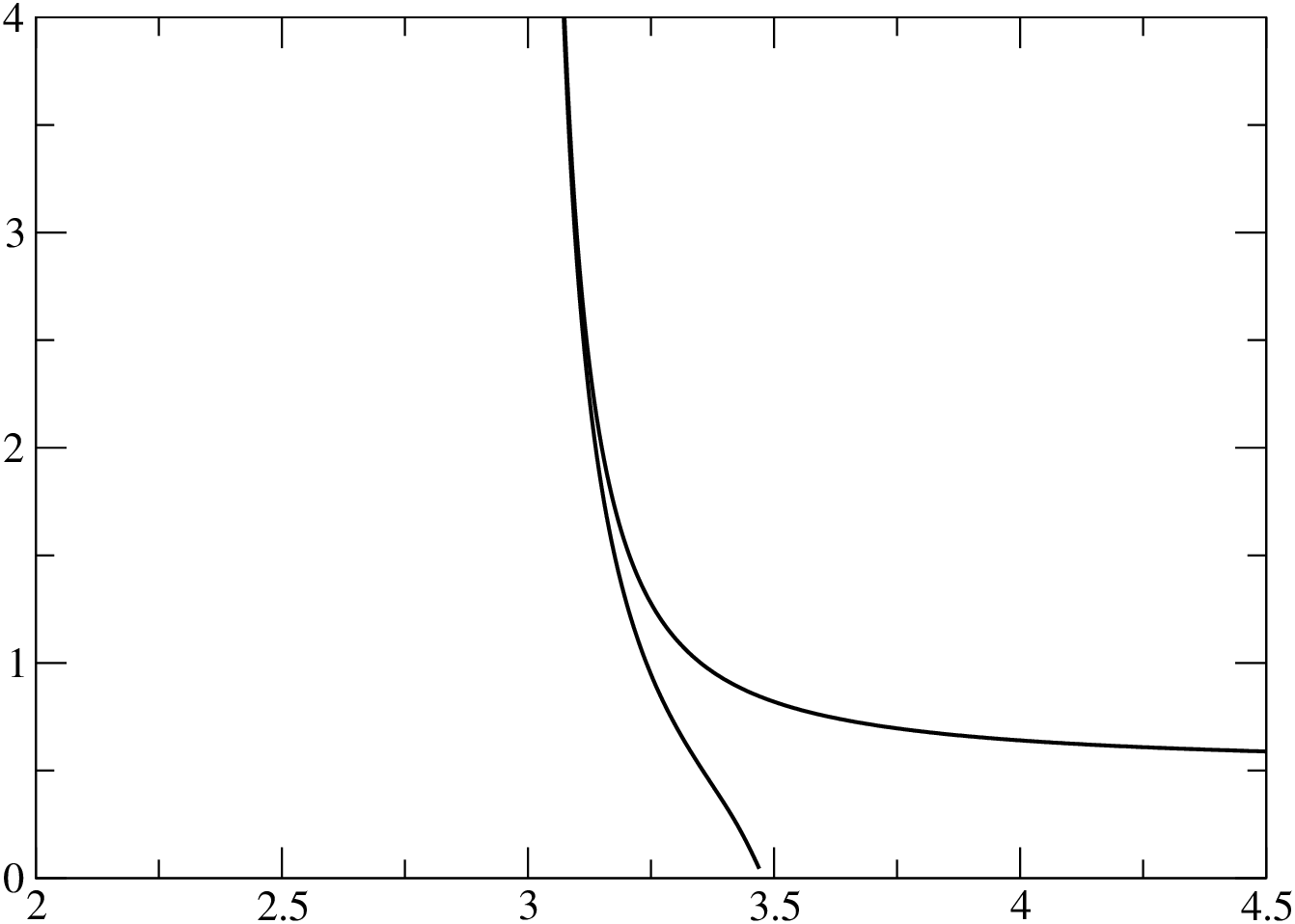}}
\put(3,60){\includegraphics[height=4.cm,width=6.6cm]{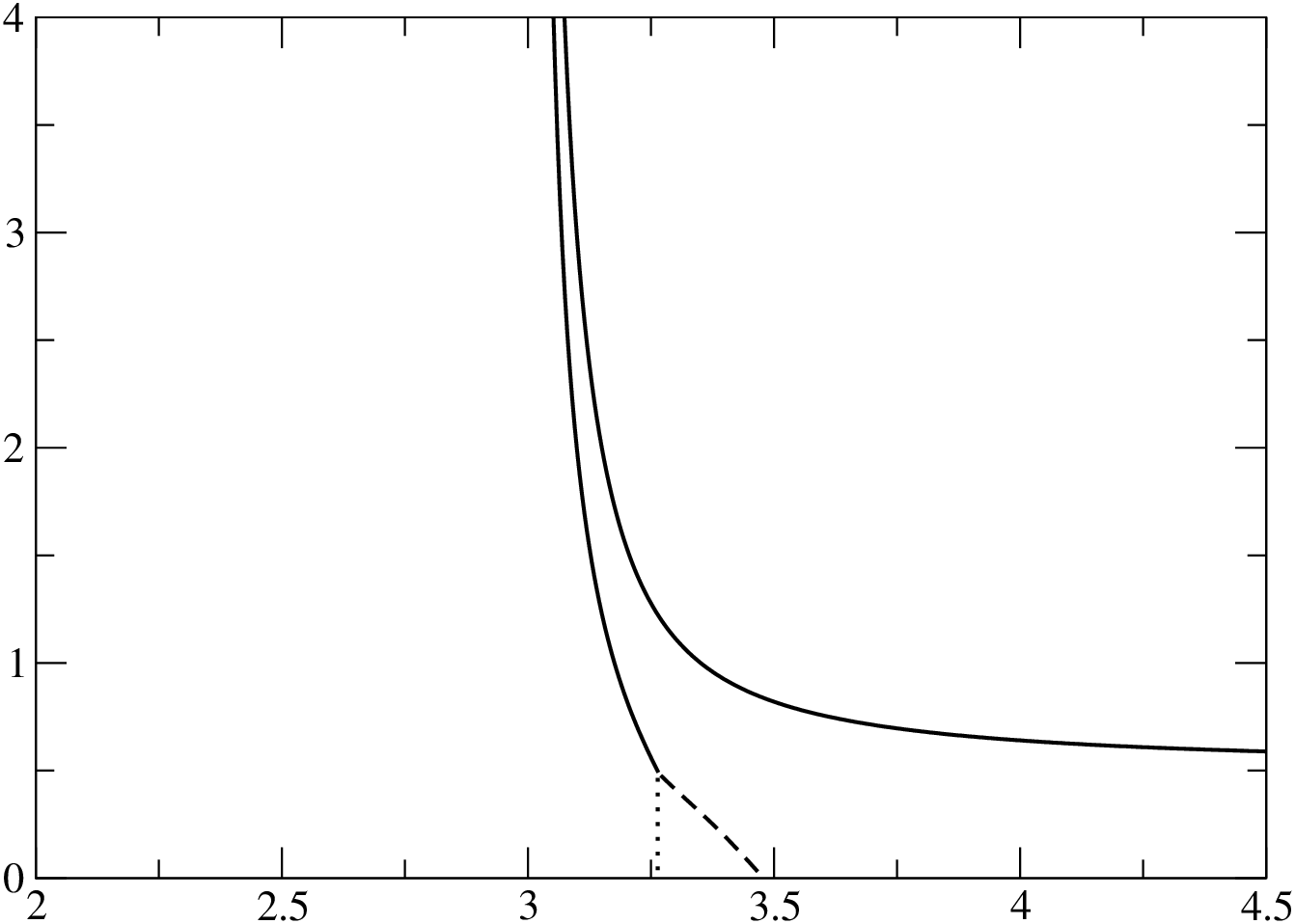}}
\put(3,10){\includegraphics[height=4.cm,width=6.6cm]{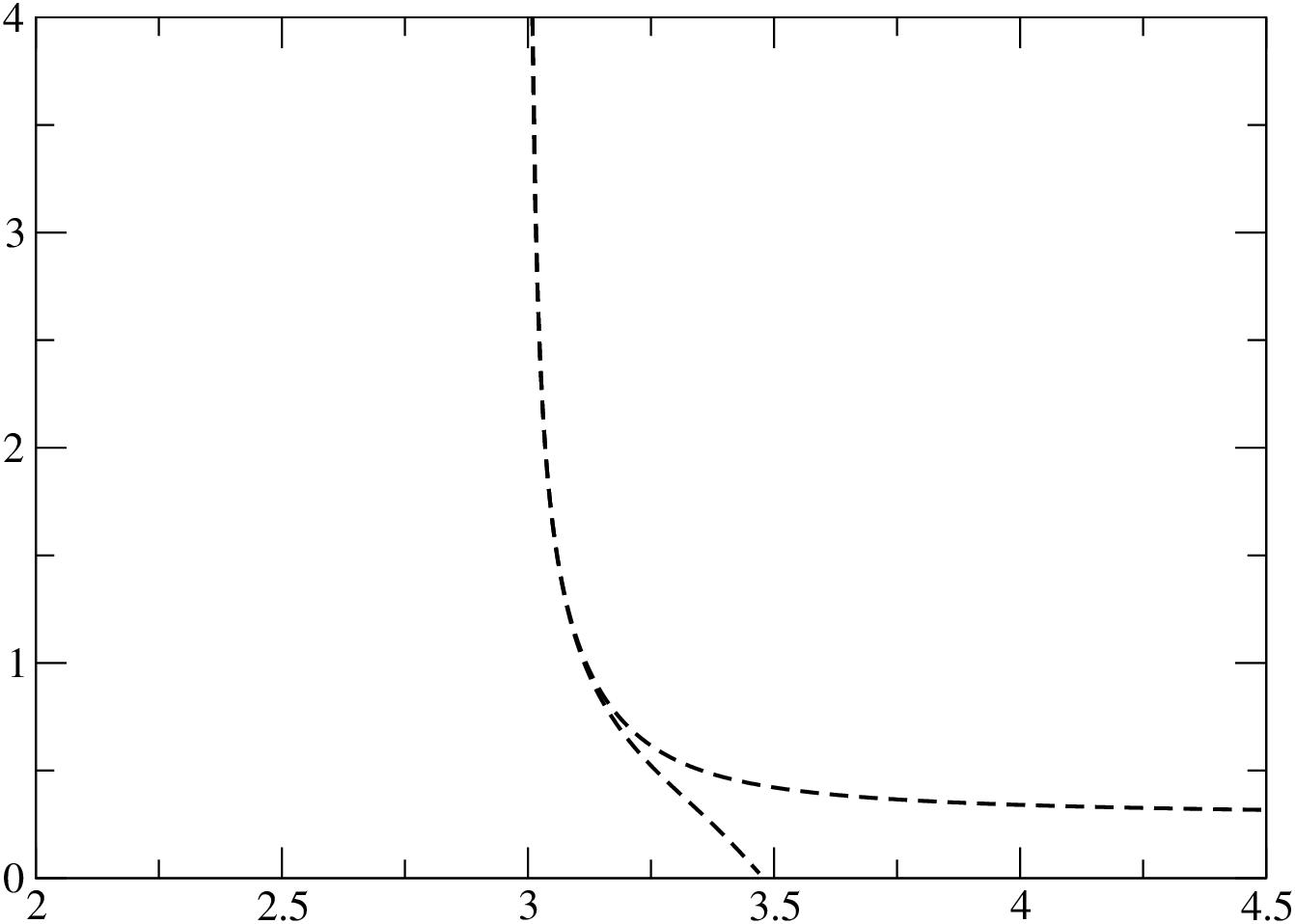}}
\put(35,5){$\gamma$}
\put(-4,30){$T/J$}\put(-4,80){$T/J$}\put(-4,130){$T/J$}
\put(40,135){P}\put(20,120){F}
\put(20,65){F}\put(40,75){P}\put(40,30){P}
\put(37,62.5){{\tiny SG}}
\put(15,15){SG}
\end{picture}
\caption{\label{fig:scalefree_bimodal}
Phase diagrams for XY small-world networks with bimodal distribution $K(\omega)=\frac12\delta[\omega-\overline{\omega}]+\frac12\delta[\omega+\overline{\omega}]$ on scale-free graphs for which $p_k\sim k^{-\gamma}$ with $k=m,\ldots,\infty$ and $\overline{\omega}=0$ (top), $\overline{\omega}=\frac14\pi$ (middle) and $\overline{\omega}=\frac12\pi$ (bottom).  We have taken the value of the lower cutoff of $p_k$ to be $m=1$. Solid lines represent P$\to$F transition whereas dashed ones represent P$\to$ SG ones (lower lines: $J_0=0$ and upper: $J_0=0.5$) and the dotted line represents F$\to$ SG. Small world effects play a significant role on the phase diagram as even for small $J_0$ values the ferromagnetic area increases substantially.}
\end{figure}
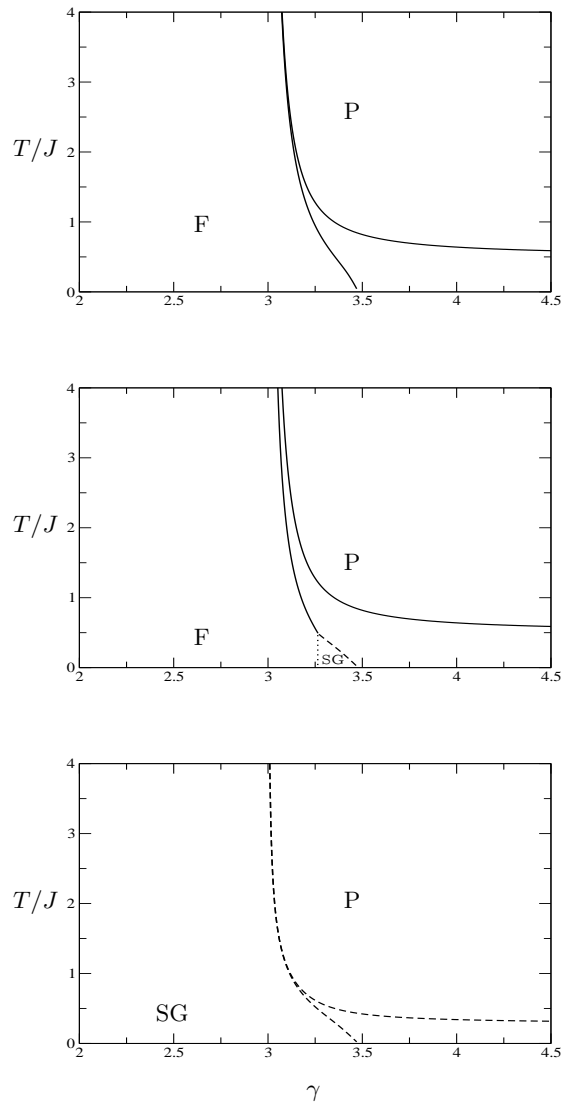
On the top panel of Fig. \ref{fig:phase_bimodal} we find the simplest possible scenario of our model \eqref{eq:hamiltonian} with $p_k=e^{-c}c^k/k!$ and $\overline{\omega}=0$. We plot the P (paramagnetic) $\to$ F (ferromagnetic) line (\ref{eq:PF}) for $J_0=0,0.5$ (with $J_0=0$ returning the results of \cite{softguzai}). In this simple case, we clearly see  the small-world effect: even for connectivity values $c\ll 1$ synchronization is possible and the F region in parameter space becomes significantly enlarged. The impact of rewiring is such that any long range connectivity value, however small (but finite), can lead to coherence at a finite temperature. A similar result based on Monte Carlo simulations has been obtained in \cite{kim,hong}.  Indicatively, even for $c=0.01$ one can find a P$\to$F transition temperature from (\ref{eq:PF}) at a finite value (around $T/J\approx 0.02$).

For $\overline{\omega}= \frac12\pi$, because of the anti-aligning forces imposed by the angular disorder, the P$\to$SG (spin-glass) transition always occurs first (i.e.\@ for higher temperatures than the P$\to$F) and synchronization is impossible. The only possible phases are paramagnetic with $m=q=0$ and spin glass with $m=0$ and $q\neq 0$. For intermediate values, e.g.\@ $\overline{\omega}=\frac14\pi$ we see that one can have all three regions in the phase diagram, although the location of the F$\to$SG line cannot be obtained from our bifurcation analysis. We have assumed that this is given by the line segment parallel to the $T/J$ axis connecting the triple point where all phases meet and $T/J=0$ \cite{Parisi1980}. This assumption is based on physical grounds (absence of re-entrance phenomena). For $\overline{\omega}=\frac14\pi$ we also see that increasing the short-range bond $J_0$ leads very quickly to an effective elimination of the SG area (for this example no SG phase appears for at least $c\leq 0.01$). Stability within the ordered phase, for parameter regions where all three phases can appear, can be found using our methodology of section \ref{sec:AT}. In the middle panel of figure \ref{fig:phase_bimodal} we show the boundary (dotted-dashed line) separating regions of $\lambda=0$ from $\lambda>0$. Above the dotted-dashed of figure \ref{fig:phase_bimodal} we have $\lambda>0$ and therefore initial conditions in the solution obtained via iteration of \eqref{eq:instanceequations}  are important, signaling the presence of ergodicity breaking. 

In figure \ref{fig:scalefree_bimodal} we plot phase diagrams for scale-free architectures where $p_k\sim k^{-\gamma}$ with $m\leq k<\infty$. We have plotted them in the $(\gamma,T/J)$-plane for different values of $J_0$, $\overline{\omega}$. As it has been noted in \cite{Leone2002} for the case of Ising spins, we find that for exponent values $\gamma\leq 3$ the system is always in a non-paramagnetic state and this is due to the divergence of the second moment $\bracket{k^2}$ at $\gamma=3$.  For spin systems with $J_0=0$ one finds that a lower cutoff $m=1$ leads  to a critical $\gamma_c$ at zero temperature which ceases to exist for $m>1$. However, for systems on small worlds ($J_0\neq 0$) we see that $m=1$ can no longer lead to such a transition as even an infinitesimally small number of long range short-cuts is sufficient to guarantee a ferromagnetic state at zero temperature.  Again, the small world effect is striking. In fig.\@ \ref{fig:scalefree_bimodal} we compare the phase diagrams for $J_0=0$ and $J_0=0.5$. We see that even a relatively small coupling strength $J_0$ guarantees coherence at finite temperatures and also  for small values of $\gamma$. As in the case of Poisson random graphs, without angular disorder ($\overline{\omega}=0$) there can only be a P$\to$F transition, whereas for $\overline{\omega}= \frac12\pi$ only a  P$\to$SG. Notice also that by increasing the lower cutoff value $m$ the area where coherence occurs is further enlarged, as expected, since the mean connectivity on the graph also grows. The dotted-line separating the SG and ferromagnetic phase is based on the Parisi-Toulouse hypothesis \cite{Parisi1980}. Due to numerical difficulties with population dynamics at low $\gamma$ we were unable to calculate the ergodicity breaking line with sufficient accuracy in this case.

\section{Conclusions}
Complex networks constitute an important area of research by virtue of the architecture through which microscopic elements interact with one another. Typically, only a vanishingly small fraction of the system, containing a finite number of other nodes, can communicate directly with any given node. Cooperativity on such networks can vary significantly depending upon parameters such as the degree distribution or a superimposed small-world structure. As such interaction structures are commonly found in a wide range of real networks in recent years they have attracted the attention of both experimentalists and theorists alike. From a theoretical standpoint the notion of `local neighborhoods' opens up new directions towards theories of finite dimensional systems.

In this paper we have adapted the techniques developed in \cite{softguzai} for the cavity method to study small-world effects in spin systems with real-valued variables defined on the unit circle. Such systems are not only significant from a physical point of view, but are also of special relevance in the context of e.g.\@ Josephson junctions \cite{wiesenfeld,granato,majhofer} and coupled oscillator systems \cite{kuramoto}. In special cases where the degree distribution is given exactly (e.g.\@ Poisson or scale-free) we have performed a bifurcation analysis and derived phase diagrams. From our results we see that superimposing a small-world structure significantly enlarges the region in parameter space where synchronization is possible; even for very small short-range coupling strengths and in the presence of disorder the system is able to synchronize. For scale-free architectures similar small-world effects are visible again for very small connectivity values. Interestingly,  for a lower cutoff of the scale free distribution $m=1$ the system always finds itself  in a synchronized state  along the zero temperature axis. This effect  is solely due to the small world architecture.
\begin{acknowledgements}
N. S. S. wishes to warmly thank RIKEN Brain Science Institute for their kind hospitality during the final stages of this work and the fund for Scientific Research, Flanders-Belgium. Meaningful and motivating discussions with A. C. C. Coolen, T. Nikoletopoulos and B. Wemmenhove are gratefully acknowledged. I. P. C. acknowledges EPSRC grant GR/R83712/01.
\end{acknowledgements}

\bibliography{XYsmall}

\end{document}